# Information and Helical Mechanism of Entropy Increase

Zou Dan Dan[1,2,*]

1. School of Electrical and Automation Engineering, East China Jiaotong University, Nanchang, 330013, China
2. School of Electrical Engineering and Automation, Wuhan University, Wuhan 430072, China
[*]Email:ddzou@hust.edu.cn

**Abstract**： The principle of entropy increase is not only the basis of statistical mechanics, but also closely related to the irreversibility of time, the origin of life, chaos and turbulence. In this paper, we first discuss the dynamic system definition of entropy from the perspective of symbol and partition of information, and propose the entropy transfer characteristics based on the set partition. By introducing the hypothesis of limited accuracy of measurement into the dynamical system with continuous phase space, two necessary mechanisms for the formation of chaos are obtained: the transfer of entropy from small scale to macro scale (i.e. the increase of local entropy) and the dissipation of macro information. The relationship between the local entropy increase and Lyapunov exponent of dynamical system is established. And then the entropy increase and abnormal dissipation mechanism in physical system are analyzed and discussed.

Clausius first introduced the concept of entropy in thermodynamics, and then Boltzmann and Planck defined the formula of entropy from microscopic statistics to describe the disorder degree of the system. Shannon extended statistical entropy to information theory in 1948[1]. And on this basis, Jaynes proposed the principle of maximum information entropy[2], which makes information theory integrate with physical statistics and can be applied to various engineering fields such as economy and society. Entropy, which is based on random probability, has become the universal existence of various disciplines.

While with the development of chaos theory, people gradually realize that many seemingly random sequences can be generated by simply determined functions, and the random phenomena in classical mechanics can be attributed to determined mechanical equations. If the information entropy system can be established from the perspective of dynamic system, it may promote the understanding and application of entropy and the micro basis of entropy increase [3,4].

In this paper, the entropy and information of dynamical system are studied by using the set theory system. Starting from the most basic concepts such as partition and the principle of symbol correspondence, the definition of information and entropy is based on the fineness of set partition [5]. And then the chaos of dynamical system and the dissipation of physical system are discussed. Based on information theory, it may provide a new perspective and path for nonlinear chaos and statistical physics.

In addition, the helix phenomenon and its chirality are widely existed in the research fields of biological DNA, composite metamaterials, turbulence and plasma discharges, which are closely related to life information and significantly affect the properties of substances. Molecular and carbon fiber devices with helical structure often have special physical properties that non helical structure does not have, such as tensile property and optical rotation [6-14]. However, these fields have been separated from each other for a long time, and unable to reveal their profound universal laws. As an attempt to bridge the information and helical mechanism, we hope to provide a reference for the theoretical basis of helix related fields.

Measurement can be attributed to the ability of resolution, which is a partition of the set composed of measurable objects. In set theory, the nonempty subsets $A_i (i = 1, 2, ..., m)$ of set A is a partition of A if:

（a） $A = \cup A_i$；

（b） if $A_i \neq A_j$, then $A_i \cap A_j = \varnothing$.

In some practical cases, the partition may have some randomness, which can be modified by the method of fuzzy logic. In set A, the measurement of the relationship between the objects is equivalent to the partition in the direct product $A \times A$.

On the other side, symbols are usually only noticed in the way of communication, and little attention is paid to their relationship with measurement. For a thermometer in a room, we don't need to use our own touch to sense the temperature in the room. Only need to read the thermometer with our eyes or be told the reading by others. The symbols used in actual measurement mainly focus on the corresponding relationship of function types. The definition of function can be expressed in the language of set as:

A and R are any two nonempty sets, and K is a positive integer, $A^K = A \times A \times \cdots \times A$, is the k-1 direct product of A with itself. According to a operation or rule (operator) F, if every element of $A^K$ is associated with exactly one element in R, then F is said to be a k-ary function from A to R, written as $F: A^K \to R$. By the following theorem, a function gives a partition of its own domain.

Theorem: For the function $F: A \to R$, F can induce a partition on the set A.

There is the same theorem for functions $F: A^k \to R$ on the set $A^K$. Therefore, the function as a special symbol, expresses a certain measurement of objects in the domain. Another thing to note is that the correspondences (or functions) are not necessary for measurement. For example, we can distinguish different colors without using language symbols to identify them. It can be divided into two processes: the partitions of objects, and the correspondences between subsets (equivalence classes) and symbols (or range).

The accuracy of the measurements depend on the fineness of the partitions. The finer the partition, the more determined the elements corresponding to the values, and the greater the amount of information expressed. Among them, binary partition divides a set into two non-empty subsets, which is the most basic method of partitions. Like the binary representation of numbers, all the partitions of a set can be classified as a combination of binary partitions.

When any element in the set A with N elements is explicitly specified, the number of binary partitions need to be imposed is $S_R = [\log_2 N]$, where [x] is the smallest integer not less than x. This function has the property of entropy. As a description of cognitive complexity, it is similar to the expression definition of thermodynamic entropy. We can also use ternary partition, eight partition or higher partition as the base. Because of the nature of logarithmic operation, the entropy of these partitions is in proportion to each other. Therefore, the entropy of set A is defined as:

$$S_R = \log_2 N . \qquad (1)$$

For a partition L: $\{A_1, A_2, ..., A_m\}$ of set A, the number of elements in each subset $A_i$ is $N_i$. When the fineness of partition needs to be measured, the average entropy can be defined as:

$$\overline{S}_{RL} = \sum_{i=1}^{m} \frac{N_i}{N} \log_2 N_i . \qquad (2)$$

On the basis of partition, the information entropy can be defined as:

$$I_L = -\sum_{i=1}^{m} \frac{N_i}{N} \log_2 \frac{N_i}{N}, \qquad (3)$$

The relationship between average entropy and information entropy is as follows

$$I_L = Log_2 N - \overline{S}_{RL} = S_R - \overline{S}_{RL} . \qquad (4)$$

It can be seen that the average entropy is inversely proportional to the information entropy. The finer the partition of set A is, the clearer the referential relationship is. The more information the partition of symbol has, and the greater the information entropy $I_L$ is, and the smaller the average entropy $\overline{S}_{RL}$ is.

Because of the composability of function, we can define another function on the set R, and the process of information transmission is equivalent to the process of symbol transformation. In the process of symbol transfer of irreversible mapping, information entropy (average information amount) can only decrease continuously. For reversible one-to-one mapping, entropy is invariant to transformation.

Therefore, for any two nonempty function $F: A \rightarrow R$ and $G: R \rightarrow L$, we have:

$$I_L \leq I_R . \qquad (5)$$

It can be seen that the above theorem is generally applicable to the ideal measurement and discrete situation. Information will only be lost in the process of transmission, and the information entropy will continue to decrease, which is a common law in the transmission and expression of language symbols.

The system consists of elements and their relations, while the dynamic system evolving with time can be regarded as the phase space composed of all the quantities related to time evolution and the time evolution function (flow mapping) defined on it. For the dynamical system (A, f), Where A is the phase space of the system and F is the time evolution function on the phase space A. If the time evolution function is equivalent to the information transfer function in the previous section, it is easy to understand that the information of the initial state will be lost with the evolution of time (or the iteration of the time evolution operator f). For the dissipative system composed of irreversible mapping, it is impossible to deduce the previous state of the system from the current state.

Taking the simplest binary dynamical system ({a, b}, f) as an example, if there are only two states a and b in the phase space of the binary dynamical system. There are three possible types of evolution functions defined by the binary system as follow.

1. Equilibrium: f (a)=a, f (b)=b;
2. Periodicity: f (a)=b, f (b)=a;
3. Convergence: f (a)=a, f (b)=a

For the system in case 1, the state will not change with time. While in case 2, the state will change circularly. Case 3 is completely different from case 1 and case 2. Its evolution function is an irreversible mapping. Regardless of the initial state, the system will eventually gather in state a, and elements a and b are in an unequal position.

Because case 3 is an irreversible mapping, when it is in state a at a certain time, it is impossible to judge whether a or b evolved at the last time, so the information is lost in this evolution, and the system is dissipative. While case 1 and 2 are all one-to-one mapping, and the current moment corresponds to the state of the previous moment one by one, and the entropy of the system will be conserved in this evolution. The evolution of ternary (and multivariate) dynamical systems also include the three types. Due to the unique correspondence property of the function, the evolution function of discrete elements does not have the process of increasing entropy.

For binary stochastic dynamical systems, due to the influence of many unmeasurable factors, it is impossible to determine the state of the next time completely according to the state of the previous time. We need to define the transition probability to describe the stochastic evolution of the system. We can define P (a | b) as the probability of the initial state a evolving to state b. similarly, we define P (b | a) respectively, P (a | a) and P (b | b), then the entropy of the system is described by the information entropy including the traditional probability. According to the principle of maximum information entropy, the distribution of transition probability of system evolution tends to maximize the information entropy under certain constraints. However, for above binary and multivariate dynamical systems without probability, entropy will only decrease, and there is no such maximum information entropy principle.

The above content mainly studies the discrete situations. However, for a determined continuous dynamical system, when the defined function f is a continuous mapping, even if f is irreversible, the information will continue to increase in many well known dynamical systems. And in many thermodynamic processes with time evolution, entropy is increased rather than decreased. In order to solve the contradiction of continuous system, it is necessary to introduce the hypothesis of finite precision for the measurement, that is, for continuous variables, the accuracy of actual measurement is always limited, and there is always measurement error. Therefore, for compressed mapping, when the distance between two points is compressed small enough, it is equivalent to two different points merging into the same point. In this mechanism, information will also be lost. On the contrary, when the mapping is expanding, the amount of information will increase. The part of information increase comes from the micro unmeasurable area, which is equivalent to the information transfer from small scale to large scale.

Considering a one-dimensional continuous dynamic system, where the phase space is a real number interval [0, 1], and the evolution function is f(x). Let the measurable accuracy of the phase space be ε, Then it divides the local interval of length *Δx* near a point x in phase space into *Δx*/ε phase lattices. According to Eq. (1), the local entropy of the initial continuous system can be defined as:

$$S = \log_2 \frac{\Delta x}{\varepsilon}, \tag{6}$$

After the application of evolution function f(x), the local distance changes, and the corresponding entropy at any point is (take the absolute value considering case of the inversion mapping)

$$S' = \log_2 \frac{|f(x+\Delta x) - f(x)|}{\varepsilon}. \tag{7}$$

The local entropy increase is

$$\Delta S = S' - S = \log_2 \frac{|f(x+\Delta x) - f(x)|}{\varepsilon} - \log_2 \frac{\Delta x}{\varepsilon} = \log_2 |\frac{f(x+\Delta x) - f(x)}{\Delta x}| \tag{8}$$

When ε and *Δx* are small enough, we have

$$\Delta S = \log_2 |\frac{df(x)}{dx}| \tag{9}$$

After the iterations of m times, the average entropy increase is

$$\Delta S = \frac{1}{m}\sum_{n=1}^{m} \log_2 \left|\frac{df(x_n)}{dx}\right| \tag{10}$$

When m tends to infinity, the average entropy increase of phase orbit is Lyapunov exponent, which is the most important index to describe the chaotic state of the system. It represents the average exponential divergence rate of two points with similar initial state in phase space after evolving with time. For compressed mapping, the volume of phase space is always in the state of contraction, the Lyapunov exponent is negative, and the information transfers from large scale to small scale region. For chaotic state, the system must have regions with positive Lyapunov exponent, and the information transfers from small scale to large scale in these regions, which makes the system entropy increase.

Folding process is also an important characteristic of chaos. From the view of information and entropy, folding will inevitably lead to the dissipation of information in the whole phase space (the same as case 3 in Figure 1). For example, for logistic mapping, the quadratic algebraic operation of $ax(1-x)$ is equivalent to stretching and refolding the real number axis of *x*, and then substituting it into the next iteration, which is similar to the baker's transformation. In this process, information loss and dissipation occur, and then small-scale information is continuously compensated. For the continuous system with period three, chaos also has this characteristic. Although there is no perfect folding in the actual physical mixing process, it can make the distance infinitely close by compression like a baker, and compress the information from a large scale to an unmeasurable small scale, thus forming the dissipation.

Therefore, the entropy increase mechanism in continuous mapping is a necessary condition for chaos. Chaos is always accompanied by the transfer of micro information to macro information. Another necessary condition is the continuous dissipation of macro information. For the system with this entropy generation mechanism, the constraints of the system (including conserved quantity, approximate conserved quantity and boundary) limit the further amplification of information in the macro level to a certain extent, thus ensuring the existence of macro dissipation mechanism. When the two mechanisms are formed in a continuous system at the same time, such as strange attractor, the phase point can keep moving away from the region, and then close to it, resulting in a series of chaotic characteristics.

The above mechanism can also explain that in statistical physics, especially in the description of a large number of nearly independent microscopic particle systems, the equilibrium of the system is in the distribution of maximum entropy. The physical collision scattering of microscopic particles is an entropy transfer process. In each collision process, the microscopic unmeasurable information (aiming distance) enlarges into the macro measurable range, the information transfers from small scale to large scale, and the measurable information of macro system increases, which leads to the entropy increase of the system. When the number of two or three body collisions increases, the macro system must reach the maximum entropy state under the constraint of conserved quantity as described in statistical physics. For the three wave interaction process in the fluid, if the wave is regarded as a particle, similar statistical methods can also be applied [8-10]. And the entropy transfer process similar to particle collision occurs.

Then infinite dimensional dynamic system will not only cause the information transfer in different spatial scales by compression and expansion, but also other mechanisms will cause the information transfer in different spatial scales. A typical example is the wave interaction of helical modes in three-dimensional incompressible fluid [6]:

$$\nabla \cdot \vec{u} = 0, \tag{11}$$

$$\partial_t \vec{u} + (\vec{u} \cdot \nabla)\vec{u} = -\frac{\nabla p}{\rho}. \tag{12}$$

Under the finite boundary of a cylinder, it can be expanded by Fourier series as

$$\partial_t \vec{u}(\vec{k}) = -i\sum_{\vec{k}'}[\vec{u}(\vec{k}+\vec{k}')\cdot \vec{k}]\vec{u}(-\vec{k}') + i\frac{\vec{k}}{k^2}\sum_{\vec{k}'}[\vec{u}(\vec{k}+\vec{k}')\cdot \vec{k}][\vec{u}(-\vec{k}')\cdot \vec{k}]. \tag{13}$$

Due to the incompressibility of neutral fluid, there is no compressible wave mode in the system, the streamline has no starting point and ending point, but only can rotate. And the equation has helical wave solution of $\vec{u} = (\sin(kz), \cos(kz), 0)$. It can be proved that the volume of phase space of fluid in wave number space is still conserved, and the wave number of different scales $\vec{k}'$ and $\vec{k}+\vec{k}'$ have three wave interaction to $\vec{k}$, forming multi-scale coupling of information. If the wave number is studied by the method of statistical mechanics, it will be found that the energy cascades (converges to small scale or large scale) and has the characteristics of dissipative system after truncating the unmeasurable small scale wave number. This mechanism is equivalent to the dissipation caused by compression, and belongs to the abnormal dissipation. In the plasma, due to the introduce of electromagnetic field and other complex mechanisms, there are more abundant helical phenomena and multi-scale dissipation mechanisms [11-14].

This abnormal dissipation mechanism can be understood by the principles of information theory and cryptography. Although the information and entropy of the initial state are not reduced, new information is introduced from the unknown factors outside the system or other unmeasurable scales inside the system. The new unknown information is like a key, and the three wave coupling with the initial information is like a cipher modulation, which hinders the decoding from the moment state (ciphertext) to the initial state (plaintext), thus making the initial information appear to be lost, and making the system have the property of mixing and dissipation.

Because all the dynamic equations of physical particles (Newton's theorem of motion, Schrodinger's equation) are reversible, the time evolution function is one-to-one mapping, and the information in the whole system is not lost but only transferred and exchanged. Therefore, the convergence mapping of case 3 for binary dynamic system is not applicable to the particle level of physical system. It can be seen that the dissipation process and time directivity of the physical system also depend on the entropy transfer mechanism of the continuous system, which is caused by the transfer of information in different spatial scales. The statistical ensemble method distributes the phase points into the phase lattices, which truncates the small-scale information, thus making the statistical and fluid theory (which including time irreversibility) a local approximation model compare to the particle level.

In summary, information entropy is defined from the view of set partition and correspondence of symbol, which forms the principle of information transmission and is applied to the phase space evolution of dynamic system. From the time evolution and transmission of local entropy, we can see that the evolution of discrete system phase space only has the behavior of equilibrium, periodicity and convergence, and the information will only gradually dissipate and decrease in the process of transmission and time evolution, without the spontaneous generation of information entropy. However, for the continuous dynamic system with limited measurement accuracy, there is a process of entropy transferring from small scale to large scale, which results in the entropy increase of macro scale. The mechanism of chaos generation in continuous system comes from the process of entropy transferring from small scale to large scale and information dissipating from large scale to small scale. The Lyapunov exponent related to chaos is the local average entropy increment of continuous mapping. For the physical system, the basic dynamic equation is reversible, the evolution function is one-to-one correspondence, and the information entropy should be conserved in the whole system. However, due to the compression and helix mechanisms of information entropy in different scales, irreversibility exists in many local approximate models with limited scales.

## Acknowledgement

We acknowledge the thoughtful talk with Prof. J. Zhu and H. Xie.## Reference

[1] C. E. Shannon. Amathematical theory of communication. Bell Sys Tech J, 1948, 27: 623～659
[2] E. T. Jaynes. Information theory and statistical mechanics. Phys Rev, 1957, 106(4): 620～630
[3] Kinoshita, T., Wenger, T. & Weiss, D. A quantum Newton's cradle. Nature 440, 900–903

(2006).

[4] Latora V , Baranger M , Rapisarda A , et al. The rate of entropy increase at the edge of chaos[J]. Physics Letters A, 1999, 273(1):97-103.

[5] D. Zou. Mathematical definition of general static systems. Chinese Journal of Systems Science, 2015, 23(04):11-13.

[6] T. D. Lee On some statistical properties of hydrodynamical and magneto-hydrodynamical fields. Quart Appl Math, 1952, 10: 69–74.

[7] R. H. Kraichnan. Helical turbulence and absolute equilibrium. J. Fluid Mech, 1973, 59: 745–752.

[8] Zhu, J.-Z. 2018, Phys. Fluids, 30, 031703

[9] Z. W. Xia, C. H. Li, D. D. Zou, et al. Helical Mode Absolute Statistical Equilibrium of Ideal Three-Dimensional Hall Magnetohydrodynamics[J].Chinese Physics Letters,2017,34(01):74-77.

[10] Chen Peng, Yang Yan, Zhu Jianzhou. Nonlinear chirality of plasma: Statistics of helices and Integrable structures [J]. Chinese Science: Physics, mechanics and astronomy, 2020,50 (04): 47-53

[11] T. Darny, E. Robert, S. Dozias, and J. M. Pouvesle, Helical Plasma Propagation of Microsecond Plasma Gun Discharges. Plasma Science, IEEE Trans. Plasma Sci. 2014, 42, 2506-2507.

[12] G. Xia , Z. Chen, Z. Yin, et al. Atmospheric Plasma Jet Relay Driven by a 40-kHz Power Supply and Its Representative Characteristics[J]. IEEE Transactions on Plasma Science, 2015, 43(5):1825-1831.

[13] D. D. Zou, X. Cao, X. P. Lu and K. Ostrikov, 2015 Phys. Plasmas 22, 103517.

[14] Zou Dandan, Cai Zhichao, Wu Peng, et al. Plasma characteristics of helical streamers induced by pulsed discharges [J]. Acta physica Sinica, 2017,66 (15): 206-211